\documentstyle[axodraw,epsf]{article}
\begin{document}

\newcommand{\symbolfootnote}{\renewcommand{\thefootnote}
        {\fnsymbol{footnote}}}
\renewcommand{\thefootnote}{\fnsymbol{footnote}}
\newcommand{\alphfootnote}
        {\setcounter{footnote}{0}
         \renewcommand{\thefootnote}{\sevenrm\alph{footnote}}}
\newcommand{\ba}{\begin{eqnarray}}
\newcommand{\ea}{\end{eqnarray}}
\newcommand{\crn}{\nonumber \\}
\newcommand{\gm}{\gamma}
\newcommand{\Gm}{\Gamma}
\newcommand{\e}{\varepsilon}
\newcommand{\trg}[3]{
\begin{picture}(35,30)(5,13)
\Line(5,5)(20,30)
\Line(35,5)(20,30)
\Line(5,5)(35,5)
\Line(5,5)(3,3)
\Line(35,5)(38,3)
\Line(20,30)(20,32)
\Text(20,0)[]{$\scriptstyle #1$}
\Text(13,25)[r]{$\scriptstyle #2$}
\Text(27,25)[l]{$\scriptstyle #3$}
\end{picture}}
\newcommand{\trgleft}[3]{
\begin{picture}(35,30)(5,13)
\Line(5,5)(20,30)
\Line(35,5)(20,30)
\Line(5,5)(35,5)
\DashLine(5,5)(-10,5){2}
\Line(-10,5)(-13,3)
\Line(35,5)(38,3)
\Line(20,30)(20,32)
\Text(-3,0)[]{$\scriptstyle -1$}
\Text(20,0)[]{$\scriptstyle #1$}
\Text(13,25)[r]{$\scriptstyle #2$}
\Text(27,25)[l]{$\scriptstyle #3$}
\end{picture}
                        }
\begin{center}
{\Large \bf  Methods to calculate scalar two-loop vertex diagrams}

\vspace{4mm}

J.~FLEISCHER\\
Fakult\"at f\"ur Physik, Universit\"at Bielefeld\\
D-33615 Bielefeld, Germany\\
~E-mail: fleischer@physik.uni-bielefeld.de.\\

\vspace{2mm}

M.~TENTYUKOV\footnote{
Supported by Bundesministerium f\"ur Forschung und Technologie
under PH/05-7BI92P 9.
}\\
Joint Institute for Nuclear Research,\\
141980 Dubna, Moscow Region, Russian Federation.\\ E-mail:
tentukov@thsun1.jinr.dubna.su

\begin{abstract}
We present a review of the Bielefeld-Dubna activities on
multiloop calculations.\\
\end{abstract}

\end{center}

   In the first contribution of the above authors in these proceedings,
we have introduced our system for the automation of evaluation of Feynman
diagrams, called DIANA (DIagram ANAlyser).
   In this contribution methods for the evaluation of scalar two-loop integrals
will be discussed.\\

\section{Expansion of three-point functions in terms of
external momenta squared}

   Taylor series expansions in terms of one external momentum
squared, $q^2$ say,
were considered in \cite{Recur}, Pad\'{e} approximants
were introduced in \cite{bft} and in Ref. \cite{ft} it was demonstrated
that this approach can be used to calculate Feynman diagrams on their
cut by analytic continuation.
The Taylor coefficients are expressed in terms of ``bubble diagrams'', 
i.e.~diagrams with external momenta equal zero, which makes their
evaluation relatively easy.
In the case under consideration
we have two independent external momenta in $d=4-2 \varepsilon$ dimensions.
The general expansion of (any loop) scalar 3-point function with its
momentum space representation $C(p_1, p_2)$ can be written as
\begin{equation}
\label{eq:exptri}
C(p_1, p_2) = \sum^\infty_{l,m,n=0} a_{lmn} (p^2_1)^l (p^2_2)^m
(p_1 p_2)^n 
\label{2.2}
\end{equation}
where the coefficients  $a_{lmn}$ are to be determined from the given diagram.

For many applications it suffices to confine to the case
$p^2_1 = p^2_2 = 0$, which is e.g. physically realized in the case of the
Higgs decay into two photons ($H \to \gamma \gamma$) with $p_1$ and $p_2$
the momenta of the photons. In this case only the coefficients $a_{00n}$ are
needed. 

\begin{figure}[h]
\centerline{\vbox{\epsfysize=45mm \epsfbox{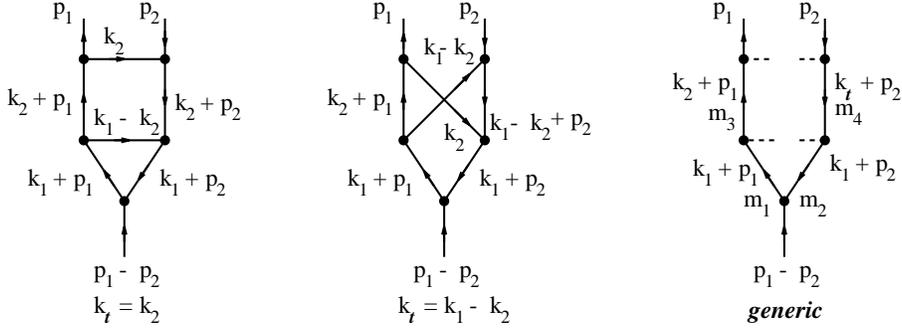}}}
\caption{\label{fig3} Planar and non-planar scalar vertex diagrams and 
their kinematics}
\end{figure}

  In the two-loop case we consider the scalar
integral ($k_3 = k_1 - k_2$, see also Fig.~\ref{fig3})
\begin{eqnarray}
\label{treug2}
\begin{array}{l}
C(m_1, \cdots, m_6; p_1, p_2) =\\
\\
\frac{1}{(i\pi^2)^2} \int
\frac{d^4 k_1 d^4 k_2}{((k_1 + p_1)^2 -m^2_1)((k_1 + p_2)^2 - m^2_2)
((k_2 + p_1)^2 - m^2_3) ((k_t + p_2)^2 - m^2_4) (k^2_2 - m^2_5)
(k^2_3 - m^2_6)}.
\end{array}
\label{2.4}
\end{eqnarray}
$k_t$ in line 4 (with mass $m_4$) depends on the topology: for the
{\it planar} diagram we have $k_t=k_2$ while for the {\it non-planar}
we have $k_t=k_3$. 



With obvious abbreviations for the scalar propagators: $c_i$ the $i^{th}$
scalar propagator of (\ref{treug2}) with $p_1=p_2=0$, we can quite generally
write for the $n^{th}$ Taylor coefficient \cite{ZiF}:
\begin{equation}
\label{e24}
(i\pi^2)^2 a_{00n} = \frac{2^n}{n+1} \int d^4 k_1 d^4 k_2 F_n \cdot
\frac{1}{c_1~ c_2~ c_3 ~ c_4~ c_5 ~ c_6}.
\end{equation}
The numerator $F_n$ can be written as 
\begin{equation}
F_n  = \sum^n_{\nu=0} c_1^{-(n-\nu)} c_3^{-\nu} \sum^n_{\nu^\prime=0}
c_2^{-(n-\nu^\prime)} c_4^{-\nu^\prime} \cdot A^n_{\nu\nu^\prime}
(k_1, k_2, k_t)\ .      \label{4.2}
\end{equation}
  For the {\it planar} diagram ($k_t=k_2$) we have
\begin{equation}
\label{hernja}
A^n_{\nu\nu^\prime} (k_1, k_2) = \sum_{\mu = \max (0,\nu + \nu^\prime -n)}^{
\left[\frac{\nu + \nu^\prime}{2}\right]} 
a^{n\mu}_{\nu\nu^\prime} (k^2_1)^{n-(\nu + \nu^\prime)+\mu}
(k^2_2)^\mu (k_1~ k_2)^{\nu + \nu^\prime - 2\mu}.
\end{equation}
The coefficients $a^{n\mu}_{\nu\nu^\prime}$ are mass independent and have
been calculated with FORM up to order ${\varepsilon}^2$ ($d=4-2\varepsilon$)
and stored for the first 30 Taylor coefficients, i.e. they are given in terms
of rational numbers. 

Finally all remaining integrals can be reduced by recursion \cite{recurrence} 
to bubble-integrals of the type
\begin{equation}
V_{\alpha\beta\gamma}(\{m\}) = \int
\frac{d^dk_1 d^dk_2}{(k_1^2-m_1^2)^{\alpha}(k_2^2-m_2^2)^{\beta}
                                           ((k_1-k_2)^2-m_3^2)^{\gamma}} ,
\label{VBs}
\end{equation}
or to factorizing one-loop integrals.The genuine two-loop bubble integrals
are reduced by means of recurrence relations to
$V_{111}(\{m\})$. This can be done numerically for the
arbitrary mass case or also analytically for special cases like e.g.
only one non-zero mass. For details see \cite{ZiF},\cite{recurrence}. 
To perform the recursion numerically,
it is important to use the multiple precision FORTRAN by D.Bailey (\cite {Bail})
since tremendeous cancellations occur in this case.

We have presented one approach for the calculation of the
Taylor expansion of Feynman diagrams in some detail, others were worked out in
Refs. \cite{davt},\cite{Tara}. The latter one is particularly suited for
programming in terms of a formulae manipulating language like FORM.

\section{The method of analytic continuation}

We assume the Taylor expansion of a scalar diagram to be given 
in the form
$C(p_1, p_2,\dots)=\sum^\infty_{m=0} a_m y^m \equiv f(y)$
and the function on the r.h.s. has a cut for $y \ge y_0$.
If the diagram has a threshold at $4m^2$ it is suggestive to introduce
$y =q^2/4m^2$ with $q^2=(p_1 - p_2)^2$ as
adequate variable with $y_0=1$.

 The method of evaluation
of the original series consists in a first step in a conformal mapping 
of the cut plane into the unit circle and secondly the reexpansion
of the function under consideration
into a power series w.r.t. the new conformal variable.
A variable often used is
\begin{equation}
\omega=\frac{1-\sqrt{1-y/y_0}}{1+\sqrt{1-y/y_0}}.
\label{omga}
\end{equation}

\begin{figure}[h]
\centerline{\vbox{\epsfysize=45mm \epsfbox{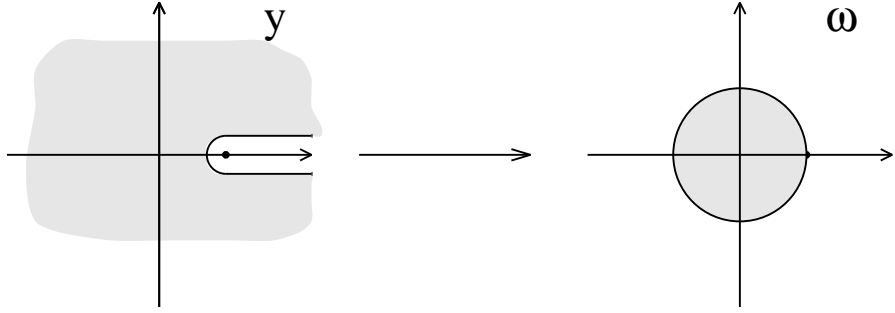}}}
\caption{\label{conf}Conformal mapping of the 
$y=q^2/4m_{\rm t}^2$-complex plane into the $\omega$-plane.}
\end{figure}

By this conformal transformation,
the $y$-plane, cut from $y_0$ to $+ \infty$, is mapped into the unit
circle (see Fig.\ref{conf}) and the cut itself is mapped on 
its boundary, the upper
semicircle corresponding to the upper side of the cut.
The origin goes into the point $\omega=0$.\\

  After conformal transformation it is suggestive to improve the
convergence of the new series w.r.t. $\omega$ by applying the 
Pad\'e method \cite{Sha},\cite{BGW}.
A convenient technique for the evaluation of Pad\'e approximations 
is the $\varepsilon$-algorithm of~\cite{Sha} which allows one
to evaluate the Pad\'e approximants recursively.

   Many interesting examples for the efficiency of this method
have been worked out in the meantime and we refer to the literatur
for massive diagrams to \cite{ft}, \cite{AIHENP}, \cite{Rheinsb} e.g.

\section{Two-loop vertex diagrams with zero thresholds}

\begin{figure}[h]
\centerline{\vbox{\epsfysize=45mm \epsfbox{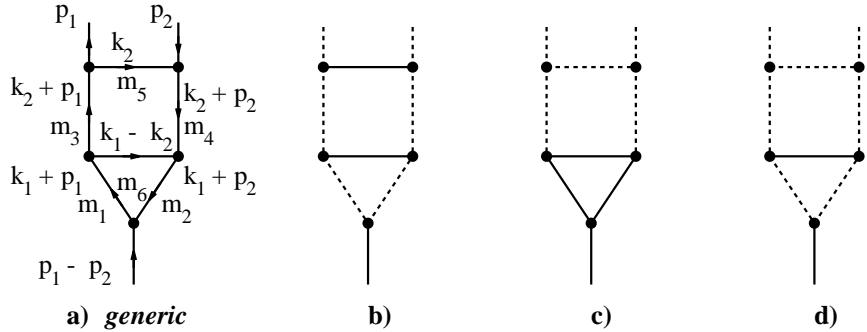}}}
\caption{\label{fig4} Planar diagrams with zero thresholds.
a)---generic;
b)---IR finite diagram; c),d)---IR divergent diagrams.}
\end{figure}

 Concerning the vertex diagrams, there are many different topologies
contributing to a 3-point function in the SM. For our
purpose of demonstra\-ting the method, we confine ourselves to the
planar case shown in Fig.~\ref{fig4}a.
Fig.~\ref{fig4}c,d presents infrared divergent diagrams.

As a typical (and very complicated case) we consider here Fig.~\ref{fig4}b
with equal non-zero masses \cite{FST}. There are two massless cuts
so that we shall have the double logarithm in the expansion.
After summing up four contributions we see that the double and
single poles in $\varepsilon$ cancel as well as the scale parameter $\mu$,
with the result $(x=q^2/M^2)$
\begin{eqnarray}
F (q^2,M^2) =
\frac{1}{M^4} \sum_{n=0}^{\infty} \sum_{j=0}^{2}
f_{jn} \ln^j(-x) x^n \nonumber \\
\label{AE22}
\equiv \frac{1}{M^4} \left\{ f_0(x) + f_1(x) \ln(-x)
+ f_2(x) \ln^2(-x)\right\} ,
\end{eqnarray}
where the $f_{jn}$ are now given in terms of rational numbers and $\zeta(2)$.
$f_2(x)$ can be summed analytically, yielding
\begin{equation}
f_2(x)= \bigl( \ln|1+x| -i\pi\theta(1+x) \bigr)^2/x^2  \,.
\end{equation}
Thus, we have to Pad\'{e} approximate $f_0$ and $f_1$ only.
Close to
the second threshold at $q^2=M^2$ the convergence is indeed excellent
(see Table \ref{padecut}) \cite{FST}. It should
be noted that for the physical application we have in mind, i.e. 
$Z \to b\bar{b}$,
this is just the case of interest. It is worthwhile to note the sharp increase
for low $q^2$ due to $\ln^2(-q^2/M^2)$.

The infrared divergent diagrams (Fig.~\ref{fig4}c,d)
have been successfully considered 
in \cite{Mainz}.

{\scriptsize
\begin{table}
\caption{Results for timelike $q^2$ for diagram \ref{fig4}b.}
\medskip
\begin{center}
\begin{tabular}{|c|ll|ll|}
\hline
$q^2/M^2$ & \phantom{0}[12/12] && \phantom{0}[15/15] & \cr
 &\phantom{0}Re & \phantom{0}Im & \phantom{0}Re & \phantom{0}Im  \cr
\hline
\phantom{0}0.05 & $+$2.948516245  & $ $20.938528 & 
$+$2.948516245  & $ $20.938528\cr
\phantom{0}0.1\phantom{0} & $-$1.108116127  & 
$ $16.04132127 & $-$1.108116127  & $ $16.04132127\cr
\phantom{0}0.5\phantom{0} & $-$4.820692281  & 
$ $\phantom{0}5.066080015 & $-$4.820692281  & $ $\phantom{0}5.066080015\cr
\phantom{0}1.0\phantom{0} & $-$3.890154  & $ $\phantom{0}1.67549787 & 
$-$3.890156  & $ $\phantom{0}1.67549788\cr
\phantom{0}1.5\phantom{0} & $-$2.904588  & $ $\phantom{0}0.42979 & 
$-$2.904581  & $ $\phantom{0}0.429778\cr
\phantom{0}2.0\phantom{0} & $-$2.18294   & $-$0.06976 & $-$2.182981  & 
$-$0.069728\cr
10.0\phantom{0} & $-$0.191   & $-$0.208 & $-$0.194   & $-$0.215\cr
\hline
\end{tabular}
\end{center}
\label{padecut}
\end{table}}

\section{The Differential Equation Method}

  We saw that to obtain the expansion
of a diagram one has to go through a rather combersome machinery.
The more coefficients are asked for, the more efforts and machine
resourses are required. Thus it is very desirable to have analytic 
expressions for expansion coefficients whenever possible. 
This can be done
with the aid of the Differential Equation Method (DEM) \cite{DEMREW}
if only one non-zero mass occurs. 
The DEM allows one to get results for massive diagrams by reducing
the problem to diagrams with simpler structure. 

  Let us introduce a graphical notation for the scalar propagators
(in euclidean space-time) 

\vspace{-5pt}\hfill\\
\hspace*{3cm}
$\frac{\displaystyle 1}{\displaystyle (q^2)^\alpha}\,=\;\;$
\begin{picture}(30,10)(5,4)
\DashLine(0,5)(30,5){2}
\Vertex(0,5){1}
\Vertex(30,5){1}
\Text(15,7)[b]{$\scriptstyle\alpha$} 
\end{picture} 
$,\qquad \frac{\displaystyle 1}{\displaystyle (q^2+m^2)^\alpha}\,=\;\;$
\begin{picture}(30,10)(5,4)
\Line(0,5)(30,5)
\Vertex(0,5){1}
\Vertex(30,5){1}
\Text(15,7)[b]{$\scriptstyle\alpha$} 
\Text(15,3)[t]{$\scriptstyle m^2$} 
\end{picture} 
\vspace{-0pt}\hfill\\
\noindent
\vspace*{2mm}
$\alpha$ and $m$ refer to the index and mass of a line.
Then one can derive the following recurrence relation for
a massive triangle \cite{DEMREW} 

\vspace{-10pt} \hfill \\
\hspace*{1cm}
\trg{\alpha_1}{\alpha_2}{\alpha_3}
$(D-2\alpha_1-\alpha_2-\alpha_3) \,=\, -2 m_1^2\alpha_1\,\,\,$
\trg{\alpha_1+1}{\alpha_2}{\alpha_3}
\vspace{20pt} \hfill \\
\vspace{-0pt}
\vspace{-20pt}\hfill \\
\hspace*{1cm}
$+\alpha_2\Biggl(\;\;$
\trg{\alpha_1-1}{\alpha_2+1}{\alpha_3}
$-\;\;\;\;\;$
\trgleft{\alpha_1}{\alpha_2+1}{\alpha_3}
$\,-(m_1^2+m_2^2)\;\;\;\;$
\trg{\alpha_1}{\alpha_2+1}{\alpha_3}
$\Biggr) + (\alpha_2\leftrightarrow\alpha_3)$
\vspace{+5pt} \hfill \\
\noindent
along with some other graphical relations (see details in \cite{DEMREW}).

  Using this technique we analysed in \cite{DEM} the class of
3-point two-loop massive graphs. As an example for the diagram 
of Fig.\ref{fig4}c we get

\vspace{-30pt}\hfill\\
\hspace*{0cm}
$(D-4)J_{\rm 5c}\,=\, 2\!\!$
\begin{picture}(35,60)(5,27)
\DashLine(5,55)(35,55){2}
\DashLine(20,30)(5,55){2}
\DashLine(20,30)(35,55){2}
\BCirc(20,17.5){12.5}
\Line(20,5)(20,3)
\Line(5,55)(3,58)
\Line(35,55)(38,58)
\Text(35,15)[l]{$\scriptstyle 2$}
\end{picture}
$\!\! -\,2\;$
\begin{picture}(35,60)(5,27)
\DashLine(5,55)(35,55){2}
\DashLine(5,30)(5,55){2}
\Line(5,30)(35,55)
\Line(20,5)(5,30)
\Line(20,5)(35,55)
\Line(20,5)(20,3)
\Line(5,55)(3,58)
\Line(35,55)(38,58)
\Text(25,15)[l]{$\scriptstyle 2$}
\end{picture}
$ -\,4m^2$
\begin{picture}(35,60)(5,27)
\DashLine(5,55)(35,55){2}
\DashLine(5,30)(5,55){2}
\DashLine(35,30)(35,55){2}
\Line(5,30)(35,30)
\Line(20,5)(5,30)
\Line(20,5)(35,30)
\Line(20,5)(20,3)
\Line(5,55)(3,58)
\Line(35,55)(38,58)
\Text(29,15)[l]{$\scriptstyle 2$}
\end{picture}
$ -\,2m^2$
\begin{picture}(35,60)(5,27)
\DashLine(5,55)(35,55){2}
\DashLine(5,30)(5,55){2}
\DashLine(35,30)(35,55){2}
\Line(5,30)(35,30)
\Line(20,5)(5,30)
\Line(20,5)(35,30)
\Line(20,5)(20,3)
\Line(5,55)(3,58)
\Line(35,55)(38,58)
\Text(20,34)[l]{$\scriptstyle 2$}
\end{picture}
\hfill (10)\\ 
\vspace{15pt}

In the r.h.s. of (10) the last two terms can be combined, 
resulting in ${\rm d}J/{\rm d}m^2$ while for the second
term we proceed in turn as

\vspace{-30pt}\hfill\\
\hspace*{2cm}
$(D-4)$
\begin{picture}(35,60)(5,27)
\DashLine(5,55)(35,55){2}
\DashLine(5,30)(5,55){2}
\Line(5,30)(35,55)
\Line(20,5)(5,30)
\Line(20,5)(35,55)
\Line(20,5)(20,3)
\Line(5,55)(3,58)
\Line(35,55)(38,58)
\Text(25,15)[l]{$\scriptstyle 2$}
\end{picture}
$=$
\begin{picture}(35,60)(5,23)
\Line(20,5)(5,40)
\Line(20,5)(35,40)
\Curve{(5,40)(20,33)(35,40)}
\DashCurve{(5,40)(20,47)(35,40)}{2}
\Line(20,5)(20,3)
\Line(5,40)(3,43)
\Line(35,40)(38,43)
\Text(28,15)[l]{$\scriptstyle 2$}
\Text(20,49)[b]{$\scriptstyle 2$}
\end{picture}
$+$
\begin{picture}(35,60)(5,23)
\Line(20,5)(5,40)
\Line(20,5)(35,40)
\Curve{(5,40)(20,33)(35,40)}
\DashCurve{(5,40)(20,47)(35,40)}{2}
\Line(20,5)(20,3)
\Line(5,40)(3,43)
\Line(35,40)(38,43)
\Text(28,15)[l]{$\scriptstyle 2$}
\Text(20,31)[t]{$\scriptstyle 2$}
\end{picture}
$-$
\begin{picture}(35,60)(5,23)
\DashLine(20,5)(5,40){2}
\DashLine(5,40)(35,40){2}
\Curve{(20,5)(25,27)(30,36)(35,40)}
\Curve{(20,5)(25,9)(30,17)(35,40)}
\Line(20,5)(20,3)
\Line(5,40)(3,43)
\Line(35,40)(38,43)
\Text(22,25)[r]{$\scriptstyle 2$}
\Text(32,15)[l]{$\scriptstyle 2$}
\end{picture}
\hfill (11)\\
\vspace{25pt}

Thus we are left with simple diagrams (these can be done
completely by Feynman parameters) and the derivative
of the initial diagram w.r.t. $m^2$. The solution of the corresponding  
differential equation in terms of a series obtained from an
integral representation reads
\begin{eqnarray}
J_{\rm 5c}&=& -\frac{\Gamma^2(1+\e)}{{(q^2)}^2 {(m^2)}^{2\e}}
\sum^{\infty}_{n=1}  
\frac{(-x)^n \Gamma^2(n)}{\Gamma(2n+1)}
\Biggl[\frac{1}{\e^2} 
~-~ \frac{1}{\e} \biggl(\ln(x) +S_1(n-1)  \biggr) - \frac{3}{2}S_2(n-1)  
\nonumber \\ &-&   \frac{15}{2}S^2_1(n-1) 
 +4S_1(n-1)S_1(2n) - \zeta(2)
 - \ln(x)S_1(n-1) + \frac{1}{2}\ln^2(x)
  \Biggr],
 \nonumber
\end{eqnarray}\\
where
$$
 S_l(n)=\sum^{n}_{1}\frac{1}{k^l}
$$\\ 
 A similar formula was obtained for the 
diagram of Fig.\ref{fig4}d (see \cite{DEM}).\\

\section{Large mass expansion versus small momentum expansion}

  If there are two or more different masses involved the 
coefficients $a_{lmn}$ in (\ref{2.2}) are not just numbers 
any more but complicated functions of mass ratios. 
If one of the masses is large, e.g. if the diagram contains
a top, one can try to perform a large mass (LM, see \cite{asymptotic}) 
expansion rather than a small momentum expansion.\\

Let us consider the diagram with $m_1=m_2=m_{top}, m_1=m_2=m_5=0$ and
$m_6=m_W$ (see e.g. \cite{FKV}). There are again three subgraphs to
be expanded. By direct evaluation we find that there are induced
poles of the order $1/\epsilon^3$ in subgraphs while in the sum
they cancel which serves as a good check. For this particular
diagram the result of the LM expansion looks like
\begin{eqnarray} \nonumber
   {\rm ~~~~~~~~~~~~~~~~dia} &=&  \sum_{n=-1}^\infty A_n
   =
   \left( b^{(0)}_n
         +b^{(1)}_n L
         +b^{(2)}_n L^2
         +b^{(3)}_n L^3 \right)
   \left(\frac{1}{m_{\rm top}^2}\right)^n\, ~~~~~~~~~~~~~~~~~~~~~~~~~~~~~~~~~(12) 
\end{eqnarray}
with $L=\log(m_{\rm top}^2/\mu^2)$ and
$b^{(i)}_n$'s being known functions of $q^2,\,m_W^2$ and $\mu^2$.\\

The considered diagram is evaluated numerically at $\mu=m_{\rm top}=180\,, m_W=80\,,
q^2=90^2$. It turns out that also here again it is extremely useful to
improve the convergence of the LM expansions by applying Pad\'e approximants.
Up to a normalization factor $\frac{1}{16\pi^2}\frac{1}{m_{top}^4}$ we
obtain for the diagram (-9.996,17.9527), taking into account terms up to
n=10 in (12). The Taylor expansion with 8 coefficients yields
a precision of appr. 10 decimals the reason for which is that $q^2$ is far 
below the threshold.
The advantage of the LM expansion, however, is that it is easier to program
and for many purposes the achieved precision is high enough.

\section{Conclusion}
  Any involved calculation in field theory nessesarily consists of
two parts: 1)automatic generation of Feynman diagrams and
source codes and 2)techniques of evaluating scalar Feynman diagrams.
In our contributions we presented both and the methods have a
standard that soon important applications can be considered.


\begin{thebibliography}{99}
\bibitem{Recur} A.I.~Davydychev and J.B.~Tausk,
{ Nucl.~Phys.},        {\bf B397} (1993) 123.
\vspace{-2.5mm}
\bibitem{bft} D.J.~Broadhurst, J.~Fleischer and O.V.~Tarasov,
{  Z.Phys.}, {\bf C 60} (1993) 287.
\vspace{-2.5mm}
\bibitem{ft}
J.~Fleischer and O.V.~Tarasov,
{  Z.Phys.}, {\bf C 64} (1994) 413.
\vspace{-2.5mm}
\bibitem{ZiF}
J.~Fleischer and O.V.~Tarasov, in proceedings of the ZiF conference on
{  Computer Algebra in Science and Engineering},
Bielefeld, 28-31 August 1994, World Scientific 1995, J. Fleischer,
J. Grabmeier, F.W.Hehl and W.K\"uchlin editors.
\vspace{-2.5mm}
\bibitem{recurrence}
K.G. Chetyrkin and F.V. Tkachov, Nucl. Phys. {\bf B192} (1981) 159;
F.V. Tkachov,  Phys. Lett. {\bf 100B} (1981) 65.
\vspace{-2.5mm}
\bibitem{Bail} D.H.~Bailey,
{\em ACM Transactions on Mathematical Software}, 19 (1993) 288.
\vspace{-2.5mm}
\bibitem{davt} A.I.~Davydychev and J.B.~Tausk,
{ Nucl.~Phys.}, {\bf B465} (1996) 507.
\vspace{-2.5mm}
\bibitem{Tara} O.V.~Tarasov,
{ Nucl.~Phys.},        {\bf B480} (1996) 397.
\vspace{-2.5mm}
\bibitem{Sha} D.~Shanks, { J.~Math.~and ~Phys.} (Cambridge, Mass.) {\bf 34} 
(1955) 1;
P.~Wynn, { Math.~Comp.} {\bf 15} (1961) 151; G.A.~Baker,
P.~Graves-Morris, Pad\'e approximants, in { Encyl.of math.and its
appl.,} Vol. {\bf 13, 14}, pp Addison-Wesley (1981).
\vspace{-2.5mm}
\bibitem{BGW} G.A.~Baker, Jr., J.L.~Gammel and J.G.~Wills,
{ J.Math.Anal.Appl.}, {\bf 2} (1961) 405;
G.A.~ Baker, Jr.,
{ Essentials of Pad\'e Approximants}, pp Academic Press (1975).
\vspace{-2.5mm}
\bibitem{DSZ} A.~Djouadi, M.~Spira and P.M.~Zerwas, {  Phys.~Lett.},
{\bf B311} (1993) 255.
\vspace{-2.5mm}
\bibitem{AIHENP}
J.Fleischer,  Proceedings of Fourth International Workshop on Software
Engineering, Artificial Intelligence and Expert System for High Energy and
Nuclear Physics (Pisa, Italy, April 3-8, 1995), p 103.
\vspace{-2.5mm}
\bibitem{Rheinsb}
J.~Fleischer and O.V.~Tarasov,
Proceedings of the 1996 Zeuthen Workshop
on Elementary Particle Theory : QCD and QED in higher Orders
(Rheinsberg, Germany, April 21 - 26, 1996).
\vspace{-2.5mm}
\bibitem{FST}
J.~Fleischer, V. Smirnov and O.V.~Tarasov,
Z. Phys. {\bf C74} (1997) 379.
\vspace{-2.5mm}
\bibitem{Mainz} J.~Fleischer et al., hep-ph/9704353, to be published in
Z. Phys. {\bf C}.
\vspace{-2.5mm}
\bibitem{DEMREW}
A.V. Kotikov, Phys.Lett. {\bf B254} (1991) 185;
ibid. {\bf B259} (1991) 314; ibid. {\bf B267} (1991) 123.
\vspace{-2.5mm}
\bibitem{DEM} J.~Fleischer, A.V.~Kotikov and O.L.~Veretin,
Bielefeld preprint BI-TP-97/26, Phys.Lett. in print; (hep-ph/9707492). 
\vspace{-2.5mm}
\bibitem{asymptotic}
F.V.~Tkachov, Preprint INR P-0332, Moscow (1983); P-0358, Moscow (1984);
~~K.G.~Chetyrkin,
Teor. Math. Phys. 75 (1988), 26; ibid 76 (1988), 207;
Preprint, MPI-PAE/PTh-13/91, Munich (1991);
~~V.A.~Smirnov,
Comm. Math. Phys. 134 (1990), 109;
{\it Renormalization and asymptotic expansions}
(Birkh\"auser, Basel, 1991).
\vspace{-2.5mm}
\bibitem{FKV}
J. Fleischer, M. Kalmykov and O. Veretin,
{\it Large Mass Expansion versus Small Momentum Expansion of
Feynman Diagrams}, Bielefeld preprint BI-TP-97/43, in preparation.
\end{thebibliography}
\end{document}